\documentclass[fp,twocolumn]{jpsj3}
\usepackage{txfonts}

\usepackage{graphicx}
\usepackage{bm}

\title{Ground state phase diagram of twisted three-leg spin tube in magnetic field}

\author{Kouki Yonaga\thanks{yona@cmpt.phys.tohoku.ac.jp} and Naokazu Shibata}
\inst{Department of Physics, Tohoku University, Sendai, 980-8578, Japan} 

\abst{
We study the ground state phase diagram of the twisted three-leg spin tube in magnetic fields by the density matrix renormalization group (DMRG) method. 
The twisted spin tube is composed of triangular unit cells and possesses strong quantum fluctuations under geometrical frustration. 
We apply the sine square deformation method to remove strong boundary 
effects and obtain smooth magnetization curves without steps of finite systems.
With the analysis of the magnetization curves and correlation functions 
we determine the ground state phase diagram consisting of 
(a) a Tomonaga-Luttinger (TL) liquid characterized by spin-$\frac{3}{2}$ Heisenberg model, (b) 3-sublattice state named UUD with 1/3 magnetization and (c) TL-liquid of massless chirality with 1/3 magnetization plateau, (d) TL-liquid of massless spin mode with or without chirality quasi long-range order. 
}

\begin{document}
\maketitle

\section{\label{sec:intro}Introduction}
Following Anderson's proposal of a spin liquid on 2D triangular 
lattice\cite{Anderson}, frustrated quantum spin systems have been 
studied to find novel quantum phenomena. 
Now it is widely believed that geometrical frustration 
leads to degenerate ground
states and gives rise to diverse low energy properties of quantum systems.
So called "chirality" defined by cross product of spins on a 
triangular lattice is a new degree of freedom emerged in such systems 
and the interplay between spin and chirality is an 
interesting research topic of frustrated systems. 
Here we apply external magnetic field and study the 
ground-state properties described by the spin and the chirality 
under the magnetic field. 
Since magnetic field introduces unidirectional anisotropy 
only in spin space, it generally suppresses spin fluctuations while
leaves the chirality unchanged that will lead to a new class of 
quantum state.

\begin{figure}[t]
 \centering
 \includegraphics[width=6cm,clip]{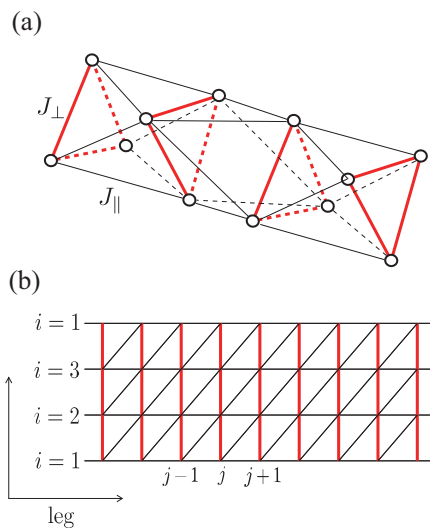}
 \caption{(Color online) (a) Structure of twisted three-leg spin tube. (b) Equivalent structure of the spin tube with the periodic boundary condition in the direction of rung. }
 \label{fig:spin_tube}
\end{figure}

In this paper we investigate the twisted three-leg spin tube 
consisting of triangular unit cells 
as a typical 1D fully frustrated quantum spin 
system (Fig.~\ref{fig:spin_tube})\cite{Nojiri_tube,Nojiri_tube2,Nojiri_tube3}. 
The ground state of this model has been studied 
mainly in several limiting cases.
In the limit of weakly interacting triangles, 
where intra-triangle interaction 
written in the red (thick) lines in Fig.~\ref{fig:spin_tube} (a) are stronger 
than the other interactions written in the black (thin) lines, 
Fouet $et\ al.$ have shown that the ground state is spin-chirality dimer state 
with translational symmetry breaking\cite{Fouet_tube,Luscher_tube,Schulz_ladder,Kawano_tube}.
In the opposite limit of strongly interacting triangles, 
they have pointed out that the effective model is a 
spin-$\frac{3}{2}$ Heisenberg model with a gapless excitation mode. 
Thus, it is expected that the ground state is characterized either 
by a spin-chirality dimer or a spin-$\frac{3}{2}$ 
quasi long-range-order, and a first order phase transition separates them
\cite{Fouet_tube,Okunishi_tube}. 

In magnetic field, Fouet $et\ al.$ found 1/3 magnetization plateau 
in the region of weakly interacting triangles\cite{Fouet_tube}.
When the intra- and inter-coupling of the triangles are comparable, 
Chen $et\ al.$ have reported that the ground state has 3-sublattice 
structure and UUD state appears in
1/3 magnetization plateau\cite{Chen_tube}. 
In this plateau, Plat $et\ al.$ have shown that
the chirality behaves as XY pseudospin in the limit of weakly interacting 
triangles\cite{Plat_tube} suggesting the chirality mode is gapless
while the spin mode is gapful in this limit. 
Although twisted three-leg spin tube is expected to have
rich phases, e.g. 3-sublattice states, two-component Tomonaga-Luttinger liquid of the spin and chirality, and 1/3 magnetization plateau, detailed ground-state 
phase diagram has not yet been clarified. 

\begin{figure}[t]
 \centering
 \includegraphics[width=6cm,clip]{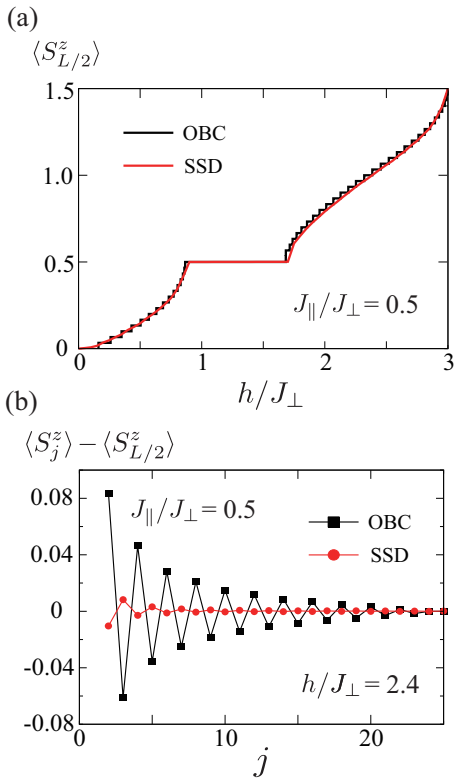}
 \caption{(Color online) Local magnetization at $J_{\parallel}/J_{\perp} = 0.5$ obtained by DMRG under the usual open boundary condition (OBC) and SSD; (a) local magnetization at the center of the system $\langle S^{z}_{L/2} \rangle = \sum_{i=1}^{3} \langle S^{z}_{i,L/2} \rangle$ and (b) site dependence of the local magnetization $\langle S^{z}_j \rangle-\langle S^{z}_{L/2} \rangle$ at $h/J_\perp=2.4$. Total number of unit triangles $L$ is $60$ and the typical truncation error in DMRG is $10^{-6}$. }
 \label{fig:magnetization_SSD}
\end{figure}

\begin{figure}[t]
 \centering
 \includegraphics[width=7.5cm,clip]{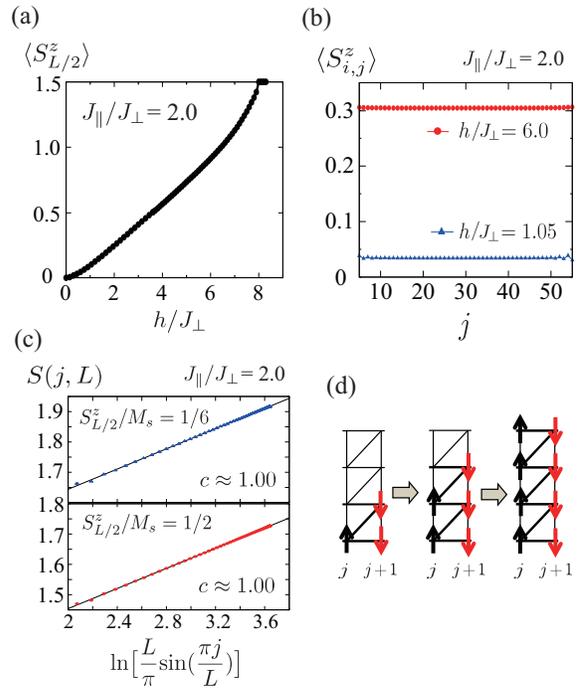}
 \caption{(Color online) Results at $J_{\parallel}/J_{\perp} = 2.0$ obtained
 by DMRG: (a) magnetization curve $\langle S^{z}_{L/2} \rangle$, (b) local
 magnetization $\langle S^{z}_{i,j} \rangle$ and (c) entanglement entropy 
 $S(j,L)$ of  block size $j$ in the $L$ site system under usual OBC. 
The upper (lower) panel in (c) shows the result for $S^{z}_{L/2}/M_{s} =
1/6$ ($S_{z}/M_{s} = 1/2$). (d) Dominant local spin structure in the 
limit of $J_{\parallel}/J_{\perp} \gg 1.0$. }
 \label{fig:J=2.0}
\end{figure}
In this paper, we apply the density matrix renormalization group (DMRG) method \cite{White_DMRG} with the sine square deformation (SSD) \cite{Nishino_SSD} and determine the ground state phase diagram of the twisted three-leg spin tube in magnetic field. The paper is organized as follows. In \S \ref{sec:model}, we define the Hamiltonian of the twisted spin tube and explain recently developed technique of SSD. In \S \ref{sec:result} and \ref{sec:discussion} our numerical results are presented and analyzed to determine the ground state phase diagram. We summarize our results in \S \ref{sec:summary}.

\section{\label{sec:model}Model and Method}
The Hamiltonian we studied here is defined by 
\begin{eqnarray}
\label{eq:H_spintube}
H    &=&  J_{\perp}\sum_{j=1}\sum_{i=1}^{3} \mbox{\boldmath $S$}_{i,j} \cdot \mbox{\boldmath $S$}_{i+1,j} \\ \nonumber
      &+& J_{\parallel}\sum_{j=1}\sum_{i=1}^{3}\mbox{\boldmath $S$}_{i,j} \cdot \mbox{\boldmath $S$}_{i,j+1} 
      + \mbox{\boldmath $S$}_{i,j} \cdot \mbox{\boldmath $S$}_{i+1,j+1}  \\ \nonumber
      &-& h\sum_{j=1}\sum_{i=1}^{3}S^{z}_{i,j} ,
\end{eqnarray}
where $\mbox{\boldmath $S$}_{i,j}$ represents spin-$\frac{1}{2}$ operator at rung $i$ and leg $j$, and $J_{\perp}\ (J_{\parallel})$ is antiferromagnetic exchange coupling of intra (inter) triangles. The last term is Zeeman energy with $h$ being the external magnetic field. 
To diagonalize this Hamiltonian we use the DMRG method, which is usually applied to open boundary conditions\ (OBC)\cite{White_DMRG}. 
However, spins at the ends of the open system cause artificial effects called "boundary effects" that sometimes make it difficult to study the bulk properties of the system. Recently, Gendiar $et\ al.$ introduced the SSD and succeed in removing boundary effects \cite{Nishino_SSD,Hikihara_SSD,Shibata_SSD}. The SSD is a kind of energy scale deformation defined as 
\begin{equation}
H_{SSD} = \sum_{j=1}^{L} f_{0}(j) h_{0}(j) + \sum_{j=1}^{L-1}f_{1}(j) h_{1}(j,j+1),
\end{equation}
where $h_{0}(j)$ is on-site term corresponding to the first and third term in Eq.(\ref{eq:H_spintube}) and $h_{1}(j,j+1)$ is the nearest neighbor interaction term such as the second term in Eq.(\ref{eq:H_spintube}). $f_{l}(j)$ is the scaling function and defined as
\begin{equation}
f_{l}(j) = {\rm sin}^{2}\big[ \frac{\pi}{L}( j + \frac{l-1}{2} ) \big].
\end{equation}

\begin{figure*}[t]
 \centering
 \includegraphics[width=14cm,clip]{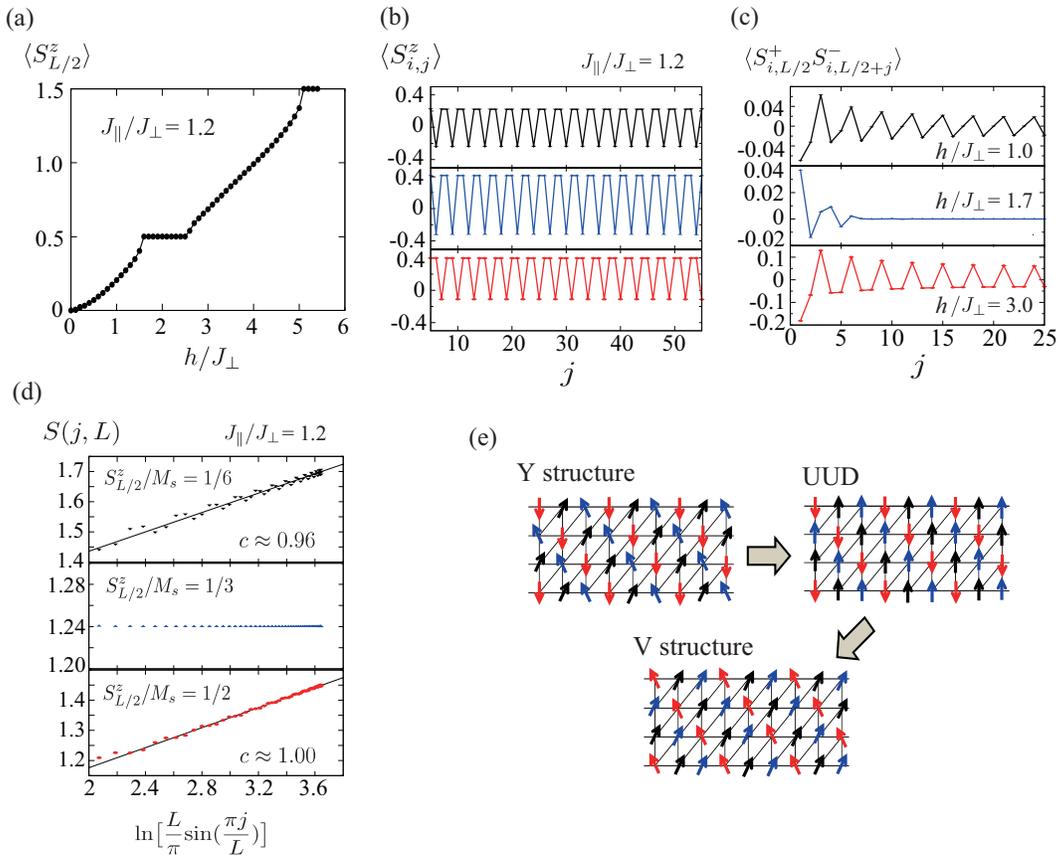}
 \caption{(Color online) Results at $J_{\parallel}/J_{\perp} = 1.2$ obtained by DMRG: (a) magnetization $\langle S^{z}_{L/2} \rangle$, (b) site dependence of local magnetization  $\langle S^{z}_{i,j} \rangle$, (c) transverse spin correlation $\langle S^{+}_{i,L/2}S^{-}_{i,L/2+j} \rangle$ and (d) entanglement entropy $S(j,L)$ at $S^{z}_{L/2}/M_{s} = 1/6$, $1/3$ and $1/2$. The upper, middle and lower panels in (b) and (c) correspond to $h/J_{\perp}=1.0, 1.7$ and $3.0$, respectively. (e) schematic spin structures of Y, UUD and V states. }
 \label{fig:J=1.2}
\end{figure*}

Since the energy scale near the edges of the system is negligibly small, the boundary effects
are efficiently suppressed around the center of the system where the scaling function is order unity. The bulk properties of the model is then evaluated around the center of the system. 
Figures \ref{fig:magnetization_SSD} (a) and (b) show the magnetization curve of $\langle S^{z}_{L/2} \rangle = \sum_{i}^{3} \langle S^{z}_{i,L/2} \rangle$ and the real space profile of the local magnetization $\langle S^{z}_{j} \rangle = \sum_{i=1}^{3} \langle S^{z}_{i,j} \rangle$. 
Without the SSD, oscillation of $\langle S^{z}_{j} \rangle$ extends into the central part of the system and the averaged magnetization is discretized by the conservation of the total $S^{z}_{j}$. In contrast, the oscillation under SSD is clearly suppressed and the magnetization curve is obtained as a smooth function of $h$.
Such removing of the boundary effects and the smooth response to the external 
field are important features of the SSD and these make it easy to find 
anomalies in the response to the external field such as magnetization plateaus.
The SSD was originally used to restore the translational invariance in 1D free fermion system and it has been shown that the ground state of critical systems such as XXZ model and (extend) Hubbard model obtained under SSD is identical to the one under the periodic boundary condition (PBC)
\cite{Katsura_SSD,Katsura_SSD2,Hikihara_SSD,Nishino_SSD2,Shibata_SSD,5}. 
In this study we use the SSD for the analysis of the ground state correlation functions and magnetization curves.

In our work, we consider the unit triangle as a single site and keep up to 1200
basis states in the DMRG calculation whose truncation error is the order of
$10^{-6} \sim 10^{-8}$. 
We show mainly the numerical results of the finite system of size $L=60$. 
Since the numerical results obtained near the edges of the system are 
meaningless in SSD, we only use the results around the center. 

\section{\label{sec:result}Results}

\begin{figure}[b]
 \centering
 \includegraphics[width=7cm,clip]{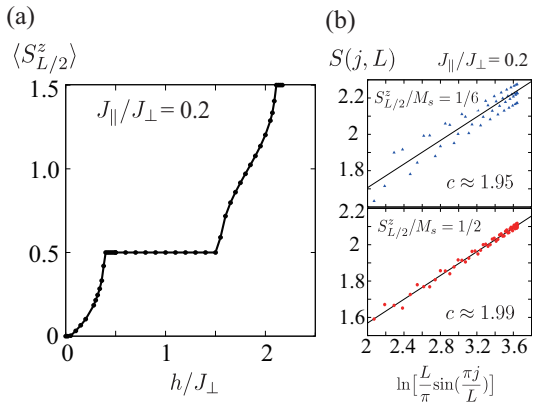}
 \caption{(Color online) DMRG results at $J_{\parallel}/J_{\perp} = 0.2$; (a) magnetization curve and (b) entanglement entropy $S(j,L)$ at $S^{z}_{L/2}/M_{s} = 1/6$ and $1/2$. } 
 \label{fig:J=0.2}
\end{figure}

\begin{figure*}[t]
 \centering
 \centering\includegraphics[width=14cm,clip]{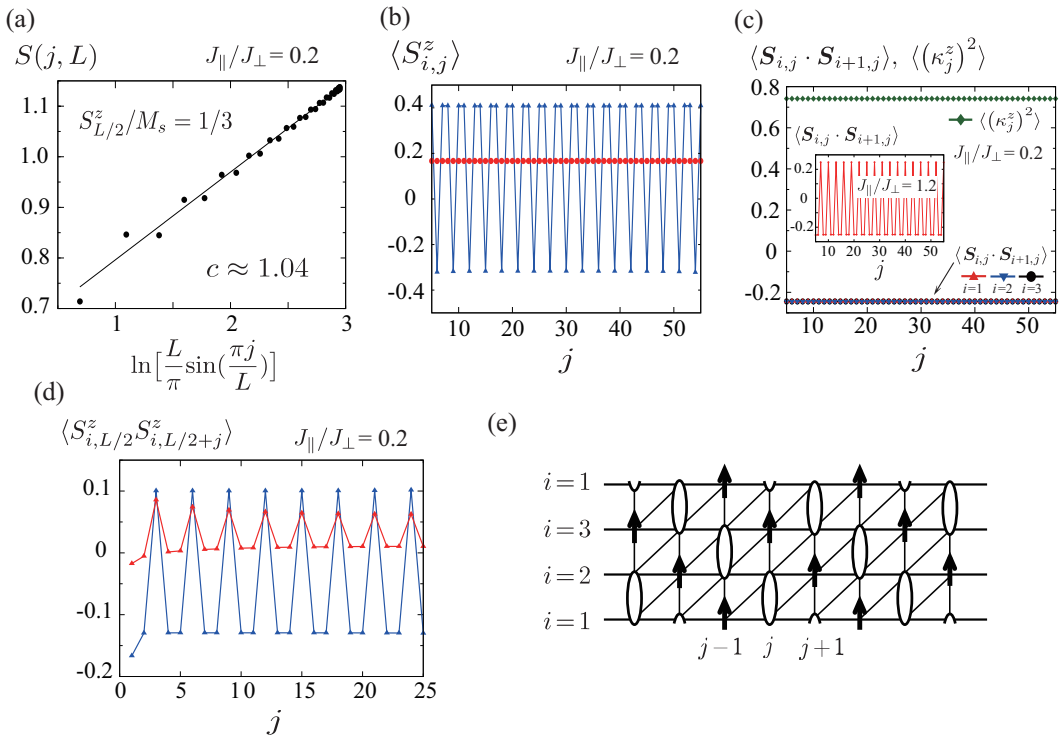}
 \caption{(Color online) DMRG results in 1/3 plateau at $J_{\parallel}/J_{\perp} = 0.2$; (a) entanglement entropy $S(j,L)$, (b) local magnetization $\langle S^{z}_{i,j} \rangle$, (c) local singlet correlation $\langle \bm{S}_{i,j}\cdot\bm{S}_{i+1,j} \rangle$ and the z-component of the local chirality $\langle \big( \kappa^{z}_{j} \big)^{2} \rangle$, (d) longitudinal spin correlation $\langle S^{z}_{i,L/2}S^{z}_{i,L/2+j} \rangle$ and (e) schematic spin structure in the 1/3 magnetization plateau at $J_{\parallel}/J_{\perp} \lesssim 1.0$. In (a), (b) and (d), we also plot the results at $J_{\parallel}/J_{\perp} = 1.2$ for comparison. The inset in (c) represents local spin correlation $\langle \bm{S}_{i,j} \cdot \bm{S}_{i+1,j} \rangle$ for UUD structure at $J_{\parallel}/J_{\perp} = 1.2$.}
 \label{fig:J=0.2plateau}
\end{figure*}

\subsection{$J_{\parallel}/J_{\perp} \gtrsim 1.0$}
We start from the case of $J_{\parallel}/J_{\perp} \gtrsim 1.0$, where
strong inter-triangle antiferromagnetic couplings ferromagnetically align 
the spins in intra-triangles as shown in Fig.\ref{fig:J=2.0} (d).  
The effective Hamiltonian is then described by spin-$\frac{3}{2}$ Heisenberg model \cite{Fouet_tube}, whose low-energy properties are characterized by a Tomonaga-Luttinger (TL) liquid. 
As shown in Fig.\ref{fig:J=2.0} (a) the magnetization at $J_{\parallel}/J_{\perp} = 2.0$ monotonically increases with the increase in magnetic field 
and the local magnetization $\langle S^{z}_{i,j} \rangle$ is site-independent. 
To confirm this gapless state we calculate the size dependence 
of the entanglement entropy which is analytically obtained for TL liquid as
\begin{equation}
S(j,L) = s_{B} {\ln} \left[ \frac{L}{\pi} {\sin}( \frac{\pi j}{L} ) \right]
\end{equation}
where $j$ and $L$ are the block and the total system size, and 
$s_{B}$ is the constant given by the central charge $c$ as 
$s_{B} = c/6\ (c/3)$ for OBC (PBC). 
Since the above formula is obtained without SSD, we calculate 
$S(j,L)$ under usual OBC without SSD.
Figure \ref{fig:J=2.0} (c) shows $S(j,L)$ at $S^{z}_{L/2}/M_{s} = 1/6$ and
$1/2$, where $M_{s}$ is the saturation magnetization per unit triangle, $3/2$.
We find the central charge $c$ is close to $1$ in both cases that 
indicates the ground state is characterized by one-component
Tomonaga-Luttinger liquid (TLL1) in agreement with the previous work by Fouet $et\ al$.

\subsection{$J_{\parallel}/J_{\perp} \approx 1.0 $}
We next consider the intermediate region $J_{\parallel}/J_{\perp} \approx 1.0$. 
Figure \ref{fig:J=1.2} (a) shows the magnetization curve at $J_{\parallel}/J_{\perp} = 1.2$ where we find clear 1/3 magnetization plateau which divides the ground state into three phases; below 1/3 magnetization plateau, on the plateau, above the plateau. 
The upper panel in Fig.\ref{fig:J=1.2} (b) shows the local magnetization
$\langle S^{z}_{i,j} \rangle$ of the ground state below the plateau. The local
magnetization $\langle S^{z}_{i,j} \rangle$ shows the presence of the 
3-sublattice structure.  
As shown in the upper panel of Fig.\ref{fig:J=1.2} (c), the transverse
correlation function $ \langle S^{+}_{i,L/2}S^{-}_{i,L/2+j} \rangle $ has
3-sites period. 
To understand this spin structure let us consider the classical limit. 
When $\langle S^{+}_{i,L/2}S^{-}_{i,L/2+j} \rangle$ is large and positive, 
$xy$ component of the two spins at $L/2$ and ($L/2+j$) sites is parallel, 
while it is almost orthogonal when
$ \langle S^{+}_{i,L/2}S^{-}_{i,L/2+j} \rangle \sim 0$, and antiparallel 
when $ \langle S^{+}_{i,L/2}S^{-}_{i,L/2+j} \rangle  < 0 $.
This simple picture indicates that the ground state below 1/3 plateau
has Y structure, which is a deformation of $120^{\circ}$ state,
where two spins are oriented upward to the $z$ direction with 
anitferromagnetic correlations in $xy$ components
while one spin is antiparallel to $z$-direction 
as shown in Fig.\ref{fig:J=1.2} (e).
Similarly we can see that the ground state above 1/3 plateau is 
characterized by V structure shown in Fig.\ref{fig:J=1.2} (e). 
As is shown in the power law decay of the correlation function
$ \langle S^{+}_{i,L/2}S^{-}_{i,L/2+j} \rangle$, 
the ground states of Y and V structures are expected to have 
massless spin excitations consistent with Mermin-Wagner theorem.
This is confirmed by the entanglement entropy $S(j,L)$ 
shown in the upper and lower panels of Fig.\ref{fig:J=1.2} (d), 
whose $j$ dependence corresponds to $c=1$ one-component TL liquid. 

In contract to the above results, 1/3 plateau state called UUD
has short range correlation as shown in the middle panel 
of Fig.\ref{fig:J=1.2} (c).
The local magnetization $\langle S^{z}_{i,j} \rangle$ 
in Fig.\ref{fig:J=1.2} (b) 
has clear 3-sublattice structure with squeezed moments 
due to quantum fluctuations.
The entanglement entropy $S(j,L)$ is independent of $j$ for large 
$j$ showing the ground state at 1/3 plateau is not critical
consistent with the short range correlation functions. 

All the 3-sublattice structures Y, UUD and V are originated from $120^{\circ}$
classical state. This was first pointed out by Chubukov in 2D triangle lattice \cite{Chubukov_tri,Miyashita_XY,Miyashita_Heisen,Griset_tri,Coletta_tri}. They have explained these 3-sublattice structures are stabilized by quantum fluctuations. 
We think the similar 3-sublattice structures in the twisted spin tube are also stabilized by the same mechanism since the unit cell of the twisted spin tube is identical to that of the 2D triangular lattice when $J_{\parallel}/J_{\perp} = 1.0 $. 

\subsection{$J_{\parallel}/J_{\perp} \lesssim 1.0$}

We finally investigate the ground state in the region of weakly interacting triangles $J_{\parallel}/J_{\perp} \lesssim 1.0$. 
Although the magnetization curve in Fig.\ref{fig:J=0.2} (a) shows 1/3 plateau at $J_{\parallel}/J_{\perp} = 0.2$, the spin structure in this plateau is different from UUD found at $J_{\parallel}/J_{\perp} \approx 1.0$ as will be shown in the following. 
In the limit of weakly interacting triangles, the effective Hamiltonian is written as
\begin{eqnarray}
\label{eq:Heff_sc}
H_{\rm eff} &=& \frac{2J_{\parallel}}{3}\sum_j \bm{S}_{j} \cdot \bm{S}_{j+1} \nonumber\\
&+& \frac{4J_{\parallel}}{3}\sum_j \bm{S}_{j} \cdot \bm{S}_{j+1}(\tau^{+}_{j} \tau^{-}_{j+1} + {\rm h.c.} )\nonumber \\
 &-& h\sum_{j=1}S^{z}_j, 
\end{eqnarray}
where $\tau^{+}_{j} (\tau^{-}_{j})$ represents the raising (lowering) operator acting on the two chirality states $|\tau_{z} = +1/2 \rangle$ and $|\tau_{z} = -1/2 \rangle$ of the unit triangle at $j$-th site, and $\bm{S}_{j}$ is spin-$\frac{1}{2}$ operator of the same unit triangle\cite{Fouet_tube,Schulz_ladder,Kawano_tube,Fuji_tube}. 
Below the magnetization plateau, the bosonization analysis shows that the spin and chirality degrees of freedoms are separated within a perturbation analysis with respect to the second term in Eq.~\ref{eq:Heff_sc}\cite{Orignac_tube,refree1,refree2,Cabra_Nleg}. To confirm this result we first calculate the entanglement entropy $S(j,L)$. 
As seen in the upper panel of Fig.~\ref{fig:J=0.2} (b), $S(j,L)$ has linear
size dependence on $\ln\left[\frac{L}{\pi}\sin{\frac{\pi j}{L}}\right]$ and
the central charge $c$ is close to $2$. This value of the central charge means the ground state below 1/3 magnetization plateau is characterized by two-component Tomonaga Luttinger liquid (TLL2)\cite{Sakai_tube,Sato_tube}. 
Similar result is also obtained above the magnetization plateau as shown in the
lower panel of Fig.~\ref{fig:J=0.2} (b). We therefore conclude 
that the elementary excitation above and below 1/3 plateau have 
two massless excitation modes for the spin and chirality.

We next see $j$-dependence of $S(j,L)$ in 1/3 plateau. In general, the
entanglement entropy becomes constant for large $j$ 
if all the excitations from the ground state have a finite gap. 
As seen in Fig.\ref{fig:J=0.2plateau} (a), $S(j,L)$ at
$J_{\parallel}/J_{\perp} = 0.2$ has linear dependence on 
$\ln\left[\frac{L}{\pi}\sin{\frac{\pi j}{L}}\right]$ with a constant 
$c$ close to 1. 
This result indicates that the elementary excitation has one massless mode. 
Since the spin excitation has a gap in the magnetization plateau, 
the low energy massless excitation of Eq.~\ref{eq:Heff_sc} is 
described by chirality\cite{Plat_tube} whose Hamiltonian is
\begin{equation}
\label{eq:Heff}
H_{\rm eff} = \frac{J_{\parallel}}{6}\sum_{j} [ 1 + 2(\tau^{+}_{j}\tau^{-}_{j+1} + {\rm h.c.}) ]. 
\end{equation}
We therefore conclude that the ground state is characterized by TLL1 
of a chirality XY model consistent with the previous work by 
Plat {\it et al.}\cite{Plat_tube}.

The local spin magnetization $\langle S_{i,j}^{z} \rangle$ of the 
above ground state is shown in Fig.~\ref{fig:J=0.2plateau} (b).
In contract to UUD state at $J_{\parallel}/J_{\perp} \approx 1.0$, 
the local magnetization at $J_{\parallel}/J_{\perp} = 0.2$ 
is uniform. The uniform ground state is
also shown in the spin correlation function 
$ \langle \bm{S}_{i,j}\cdot \bm{S}_{i+1,j} \rangle$ 
in the same triangle and the local chirality expectation value 
$\langle \big( \kappa^{z}_{j} \big)^{2} \rangle$, where $\kappa^{z}$ is given by
\begin{equation}
\kappa^{z}_{j} = \sum_{i=1}^{3} \big( \bm{S}_{i,j} \times \bm{S}_{i+1,j} )^{z}. 
\end{equation}
As shown in Fig.~\ref{fig:J=0.2plateau} (c), $\langle \bm{S}_{i,j} \cdot \bm{S}_{i+1,j} \rangle \approx -0.25$ and $\langle \big( \kappa^{z}_{j} \big)^{2} \rangle \approx 0.75$. 
These results show the coexistence of a spin singlet and chirality 
in the unit triangle.
This is contrasted with the results at $J_{\parallel}/J_{\perp} = 1.2$ 
where UUD structure appears with small $\langle \big( \kappa^{z}\big)^{2} \rangle$ and one positive and two negative local spin correlations, 
$\langle S^{z}_{i,L/2}S^{z}_{i,L/2+j} \rangle$, within a triangle. 
This difference in the local correlation makes clear difference in the long
range spin correlation $\langle S^{z}_{i,L/2}S^{z}_{i,L/2+j} \rangle$ 
shown in Fig.~\ref{fig:J=0.2plateau} (d) that is positive or 
nearly 0 at $J_{\parallel}/J_{\perp} = 0.2$,
while it has a structure of 3-site period with one positive and two negative
at $J_{\parallel}/J_{\perp} = 1.2$.
We illustrate the schematic diagram of the ground state at 
$J_{\parallel}/J_{\perp} = 0.2$ in Fig.\ref{fig:J=0.2plateau} (e). 
The ground state is a superposition of one singlet and one polarized spin 
like 0.5-0-0 in the direction of both rung and leg. 
Since each polarized spin is connected to spin singlet states, 
there is no penalty in the exchange energy among them. 
Therefor $\langle S^{z}_{L/2}S^{z}_{L/2+j} \rangle$ has 3-site period while
$\langle S^{z}_{j} \rangle$ is uniform. 
This is caused by strong geometrical frustration 
and a unique characteristic of the twisted spin tube. 
 
\begin{figure}[t]
 \centering
 \includegraphics[width=7cm,clip]{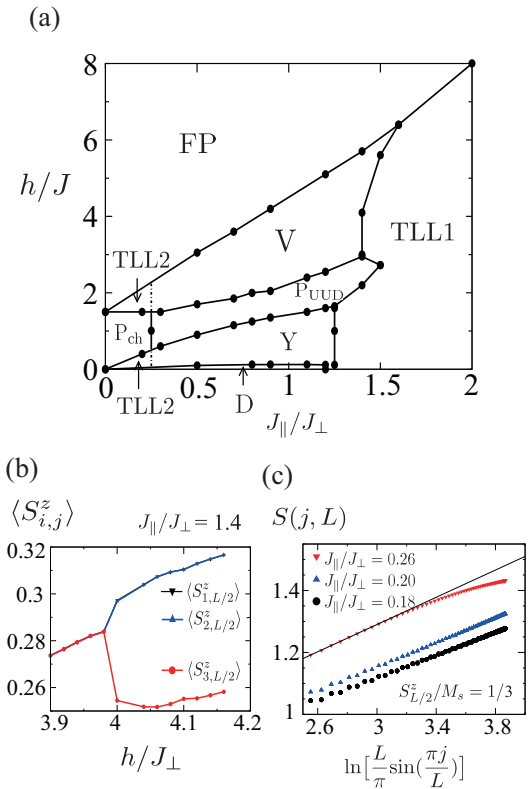}
 \caption{(Color online) (a) Ground state phase diagram of the 
twisted spin tube. 
The labels D, TLL1, TLL2 represent the dimer state, one- and 
two-component Tomonaga-Luttinger liquid, respectively. 
The labels Y and V represent Y, V states. ${\rm P_{\rm UUD} }$ 
and ${\rm P_{\rm ch}}$ corresponds to 1/3 magnetization plateau 
with UUD structure and 1/3 plateau with TLL1 of chirality, respectively. FP is fully polarized state.
(b) local magnetization $\langle S^{z}_{i,L/2} \rangle$ at 
$J_{\parallel}/J_{\perp} = 1.4$ as a function of $h/J_{\perp}$. 
(c) entanglement entropy $S(j,L)$ at $S^{z}_{L/2}/M_{s} = 1/3$. }
 \label{fig:phase_diagram}
\end{figure}

\section{\label{sec:discussion}Discussion}
The phase boundaries of the ground state obtained in each region
$J_{\parallel}/J_{\perp} \gtrsim 1.0$, $J_{\parallel}/J_{\perp} \approx 1.0$
and $J_{\parallel}/J_{\perp} \lesssim 1.0$ are presented in the phase diagram
shown in Fig.\ref{fig:phase_diagram} (a).  
The boundary is determined by the appearance or the change of 3 sublattice
structure and the central charge of TL liquids.  
For the boundary between V and TL liquid states in the region of
$J_{\parallel}/J_{\perp} > 1.0$, we have used the appearance of the 3
sublattice structure as a signal of the transition. This is clearly shown in
the local spin polarization $\langle S^{z}_{i,L/2} \rangle$ at
$J_{\parallel}/J_{\perp} = 1.4$.  
As shown in Fig.\ref{fig:phase_diagram} (b), $\langle S^{z}_{i,L/2} \rangle$
splits into two values at $h \approx 4$ which corresponds to the change 
from spin-$\frac{3}{2}$ TLL1 to V state.  
The phase diagram around the plateau state is rather complicated. 
Since the corresponding 2D triangular lattice does not have 1/3 
plateau in the region of $J_{\parallel}/J_{\perp} \lesssim 0.5 $\cite{Coletta_tri},
the appearance of the magnetization plateau with 
massless mode of chirality is a unique feature of the 
twisted spin tube. 
Above and below the magnetic plateau, the elementary excitations at
$J_{\parallel}/J_{\perp} \lesssim 0.3$ are characterized by the 
central charge $c \approx 2$ and are different from 3-sublattice 
state and TLL1.

Figure \ref{fig:phase_diagram} (c) shows $S(j,L)$ for different
$J_{\parallel}/J_{\perp}$ in 1/3 magnetization plateau. It is seen that $S(j,L)$
has linear size dependence on 
$\ln\left[\frac{L}{\pi}\sin{\frac{\pi j}{L}}\right]$
at $J_{\parallel}/J_{\perp} = 0.18$ and $0.20$, while it dose not 
at $J_{\parallel}/J_{\perp} = 0.26$. 
Thus, TLL1 of the chirality with uniform magnetization extends 
to $J_{\parallel}/J_{\perp} \approx 0.26$. 
We expect the transition from TLL1 of the chirality to
UUD state is second order because we find no signal of
level crossing, but it is difficult to confirm this point
within a finite system.
Detailed finite size scaling analysis is needed
to clarify the nature of the transition. 

\section{\label{sec:summary}Summary}
We have studied the ground state of the twisted three-leg spin tube in
magnetic field by the DMRG with the SSD and identified various 
phases such as TLL1, TLL2, 3-sublattice state and 1/3 
magnetization plateau. 
In particular we have confirmed that for $J_{\parallel}/J_{\perp} \lesssim 0.3$
the ground state in the 1/3 plateau has low energy 
excitations described by massless mode of chirality.
The coexistence of the spin and chirality degrees of freedom
and the interplay between them are characteristic features of
twisted three-leg spin tube and the origin of its diverse 
ground states.

\section*{Acknowledgment}
K.\ Y. would like to thank Y.\ Fuji for helpful advice.This work was supported by Grants-in-Aid for Scientific Research (No. 26400344) from MEXT Japan.



\begin{thebibliography}{99}
\bibitem{Anderson}
P.~W.~Anderson, Mater. Res. Bull. {\bf 8} (1973) 153.

\bibitem{Nojiri_tube}
J.~Schnack, H.~Nojiri, P.~K$\ddot{\rm o}$gerler, G.~J.~T.~Cooper and L.~Cronin, Phys. Rev. B {\bf 70} (2004) 1774420.
\bibitem{Nojiri_tube2}
N.~B.~Ivanov, J.~Schnack, R.~Schnalle, J.~Richter, P.~K$\ddot{\rm o}$gerler, G.~N.~Newton, L.~Cronin, Y.~Oshima and H.~Nojiri, Phy. Rev. Lett. {\bf 105} (2010) 037206.
\bibitem{Nojiri_tube3}
Y.~Furukawa, Y.~Sumida, K.~Kumagai, F.~Borsa, H.~Nojiri, Y.~Shimizu, H.~Amitsuka, K.~Tenya, P.~K$\ddot{\rm o}$gerler and L.~Cronin, J. Conf. Ser. {\bf 320} (2011) 012047.

\bibitem{Fouet_tube}
J.-B.~Fouet, A. L$\ddot{\rm a}$uchli, S. Pilgram, R. M. Noack and F. Mila, Phys. Rev. B {\bf 73} (2006) 014409.
\bibitem{Luscher_tube}
A.~L$\ddot{\rm u}$scher, R.~M.~Noack, G.~Misguich, V.~N.~Kotov and F.~Mila, Phys. Rev. B {\bf 70} (2004) 060405(R).
\bibitem{Schulz_ladder}
H.~J.~Schulz, cond-mat/9605075.
\bibitem{Kawano_tube}
K.~Kawano and M.~Takahashi, J. Phys. Soc. Jpn. {\bf 66} (1997) 4001.
\bibitem{Okunishi_tube}
K.~Okunishi, S.~Yoshikawa, T.~Sakai and S.~Miyashita, Int. J. Mod. Phys. C {\bf 29} (2009) 1423.
\bibitem{Chen_tube}
Ru Chen, Hyejin Ju, Hong-Chen Jiang, O.~A.~Starykh and L.~Balents, Phys. Rev. B {\bf 87} (2013) 165123.
\bibitem{Plat_tube}
X.~Plat, S.~Capponi and P.~Pujol, Phys. Rev. B {\bf 85} (2012) 174423.

\bibitem{White_DMRG}
S.~R.~White, Phys. Rev. Lett. {\bf 69} (1992) 2863; S.R. White, Phys. Rev. B {\bf 48} (1993) 10345.
\bibitem{Nishino_SSD}
A.~Gendiar, R.~Krcmar and T.~Nishino, Prog. Theor. Phys. {\bf 122} (2009) 953 {\bf 123} (2010) 393.
\bibitem{Hikihara_SSD}
T.~Hikihara and T.~Nishino, Phys. Rev. {\bf 83} (2011) 060414(R). 
\bibitem{Shibata_SSD}
C.~Hotta and N.~Shibata, Phys. Rev. B {\bf 86} (2012) 041108(R).
\bibitem{Katsura_SSD}
H.~Katsura, J. Phys. A, Math. Theor. {\bf 44} (2011) 252001.
\bibitem{Katsura_SSD2}
H.~Katsura, J. Phys. A, Math. Theor. {\bf 45} (2012) 115003.
\bibitem{Nishino_SSD2}
A.~Gendiar, M.~Dani$\check{s}$ka, Y.~Lee and T.~Nishino, Phys. Rev. A {\bf 83} (2011) 0522118.
\bibitem{5} 
N.~Shibata and C.~Hotta, Phys. Rev. B {\bf 84} (2011) 115116.

\bibitem{Chubukov_tri}
A.~V.~Chubukov and D.~I.~Golosov, J. Phys. Condens. Matter {\bf 3} (1991) 69.
\bibitem{Miyashita_XY}
S.~Miyashita and H.~Shiba, J. Phys. Soc. Jpn. {\bf 53} (1984) 1145.
\bibitem{Miyashita_Heisen}
H.~Kawamura and S.~Miyashita, J. Phys. Soc. Jpn. {\bf 53} (1984) 9.
\bibitem{Griset_tri}
C.~Griest, S.~Head, J.~Alicea and O.~A.~Starykh, Phys. Rev. B {\bf 84} (2011) 245108.
\bibitem{Coletta_tri}
T.~Coletta, M.~E.~Zhitomirsky and F.~Mila, Phys.~Rev. B {\bf 87} (2013) 060407(R).

\bibitem{Fuji_tube}
Y.~Fuji, S.~Nishimoto, H.~Nakada and M.~Oshikawa, Phys. Rev. B {\bf 89} (2014) 054425.

\bibitem{Orignac_tube}
E.~Orignac, R.~Citro and N.~Andrei, Phys. Rev. B {\bf 61} (2000) 11533.
\bibitem{refree1}
R.~Citro, E~Orignac, N~Andrei and S~Qin, J. Phys. Condens. Matter {\bf 12} (2000) 3041.
\bibitem{refree2}
Kunj~Tandon, Siddhartha~Lal, Swapan~K.~Pati, S.~Ramasesha and Diptiman~Sen, Phys. Rev. B {\bf 59} 395.
\bibitem{Cabra_Nleg}
D.~C. Cabra, A.~Honecker and P.~Pujol, Phys. Rev. B {\bf 58} (1998) 6241.

\bibitem{Sakai_tube}
M.~Sato and T.~Sakai, Phys. Rev. B {\bf 75} (2007) 014411.
\bibitem{Sato_tube}
M.~Sato, Phys. Rev. B {\bf 75} (2007) 174407.


\end{thebibliography}
\end{document}